\documentclass[prb,twocolumn,showpacs]{revtex4} 
\usepackage{graphicx}
\usepackage{amssymb,amsmath}
\begin{document}

\title{Renormalized mean-field $t$-$J$ model of  high-$\text{T}_c$  superconductivity: comparison with experiment}    

\author{Jakub J\c{e}drak}
\email{jedrak@th.if.uj.edu.pl}
\author{Jozef Spa\l ek}
\email{ufspalek@if.uj.edu.pl}
\affiliation{Marian Smoluchowski Institute of Physics, Jagiellonian University, Reymonta 4, PL-30059 Krak\'ow, Poland}

\date{\today}

\begin{abstract}
Using an advanced version of the renormalized mean-field theory (RMFT)  for the $t$-$J$ model, we examine spin-singlet superconducting (SC) state of $d_{x^2 - y^2}$-symmetry. Overall doping dependence of the 
SC gap magnitude is in good agreement with  experimental results  for  $\text{Bi}_{2}\text{Sr}_{2}\text{Ca}  \text{Cu}_2 \text{O}_{8 + \delta}$ (BSCCO) and  $\text{La}_{2-x}\text{Sr}_{x}\text{Cu} \text{O}_{4}$  (LSCO) compounds at the optimal doping and in the overdoped regime. We also calculate  the dispersion  relation for the Bogoliubov quasiparticles and compare our findings both with the angle resolved photoemission data for the cuprates, as well as with the  variational Monte Carlo and other  mean-field studies. Within the method proposed by Fukushima [cf. Phys. Rev. B \textbf{78}, 115105 (2008)], we analyze different forms of the $t$-$J$ Hamiltonian, i.e. modifications caused by the form  of exchange interaction,  and by the  presence of three-site terms. It is shown  that although  the former has a small influence,  the  latter suppresses  strongly the superconductivity. We also analyze the temperature dependence of the gap magnitude and compare the results with those of the recently introduced finite-temperature renormalized mean-field theory (TRMFT) of Wang et al. [cf. Phys. Rev. B \textbf{82}, 125105 (2010)].
\end{abstract}

\maketitle


\section{\label{sec:1} Introduction}
 One of the most characteristic features  of  high temperature superconductivity (SC) is that  upon  hole doping, with the hole concentration $x \gtrsim  0.05$, a generic antiferromagnetic Mott insulating state of e.g. $\text{La}_{2}\text{Cu} \text{O}_{4}$ \cite{Schriefer Handbook, Vaknin} transforms into SC state. The latter, in turn, after reaching a maximal transition temperature at $x\approx 0.15$, disappears at the upper critical concentration $x_c  \gtrsim 0.25 - 0.35$, depending on the system. \cite{Keimer, Nakano exp} The last property is particularly surprising, since in the overdoped regime  $x\gtrsim0.15$ the system evolves gradually from a non-Fermi liquid into a quantum liquid that can be regarded as an unconventional Fermi liquid. \cite{Fujimori} The appearance of $x_c$ may speak in favor of real-space type of pairing, as the increased hole doping reduces the pairing correlations in real space. 
To describe the above features the $t$-$J$ model  is  often invoked, \cite{Spalek PRB, Spalek Acta, Zhang Rice} and the kinetic exchange interaction is claimed to induce both antiferromagnetism and the superconductivity. \cite{Lee} 

The basic question is whether within   the  renormalized mean-field theory (RMFT) \cite{ZGRS, Edegger} we  can reproduce at least some of the above properties in a semiquantitative manner, \cite{Edegger, Plain Vanilla, Edegger Anderson Lett} as advocated strongly.  \cite{Plain Vanilla, Edegger Anderson Lett} 
On the other hand, variational Monte Carlo  (VMC)  methods, within which one treats exactly the double  occupancy exclusion, is known \cite{Paramekanti, Sorella Lett, Shih Lett, Ivanov, Spanu B, Pathak} to provide a semiquantitative description of the SC correlated state. Hence it is often regarded as being superior to any mean-field (MF) treatment. 
However, a proper MF approach would have important advantages over VMC. First, its results are not limited to small clusters. Second, it can offer an analytic insight into the physical contents of the model.  Third, it allows for a detailed comparison with experiment or, strictly speaking, its critical assessment, as detailed below. 

Numerous attempts to improve RMFT have been made, \cite{The ladder of Sigrist, Ogata Himeda, Fukushima, Wang F C Zhang} in order to take into account also intersite correlations.  Here we show that within the MF renormalization scheme proposed recently by Fukushima, \cite{Fukushima} supplemented with the maximum-entropy based \cite{Jaynes} self-consistent variational approach, \cite{JJJS PRB,  JJJS_arx_0, SGA} we can produce, among others, the results that are competitive to  those of VMC. Specifically, the upper critical concentration $x_c$, as well as the principal features of the excitation spectrum, are shown to agree quite well with experiment in the overdoped regime. Finally, motivated by a recent paper,\cite{TRMFT} we comment on the extension of the present treatment to finite temperatures. We study a behavior of the renormalized gap magnitude as a function of the temperature, which is shown to be in quite good agreement with the classic result of the BCS theory.


\section{\label{sec:2} Model and method}  
We start from $t$-$J$ model \cite{Spalek PRB, Spalek Acta, Zhang Rice, Lee}  is expressed by the following  Hamiltonian  
\begin{equation}
 \hat{H}_{tJ} = \hat{P} \Big(\hat{H}_{t} +  \sum_{\langle i j \rangle} J_{ij}~ \big( \hat{\mathbf{S}}_{i}\cdot \hat{\mathbf{S}}_{j} - \frac{c_1}{4}\hat{\nu}_{i}\hat{\nu}_{j}\big)  + c_2  \hat{H}_{3}\Big) \hat{P}.
\label{t-J exact complete}
\end{equation}
The first term, $\hat{H}_{t}= \sum_{ij\sigma} t_{ij} c_{i \sigma}^{\dag} c_{j \sigma}$ is the kinetic energy  part, the second expresses the kinetic exchange, and the third the three-site terms. $\sum_{\langle i j \rangle}$ means the summation pair of sites $\langle i,j\rangle$ (bonds). The Gutzwiller projector $\hat{P} = \prod_{i}(1-\hat{n}_{i \uparrow}\hat{n}_{i \downarrow} )$ eliminates double occupancies in real space. Also, explicitly
\begin{equation}
\hat{H}_{3} =   \sum_{i j k \sigma}  \frac{t_{ij} t_{jk}}{U} \left(  b_{i \sigma}^{\dag}  \hat{S}^{ \bar{\sigma}}_{j} b_{k \bar{\sigma}} -  b_{i \sigma}^{\dag}  \hat{\nu}_{j \bar{\sigma}}  b_{k \sigma} \right), 
\label{t-J t-s-t}
\end{equation}
 with the projected fermion operators defined as: $b_{i \sigma} \equiv (1-\hat{n}_{i \bar{\sigma}}) c_{i \sigma} $, $\hat{\nu}_{i \sigma} \equiv (1-\hat{n}_{i \bar{\sigma}}) \hat{n}_{i \sigma} $, $\hat{\nu}_{i} \equiv \sum_{\sigma} \hat{\nu}_{i \sigma}$, and $\hat{S}^{  \sigma }_{i} \equiv b_{i \sigma}^{\dag} b_{i \bar{\sigma}} $. Here the standard fermion creation (annihilation) operators are $c_{i \sigma}^{\dag}$ ($c_{i \sigma}$) and $ \hat{n}_{i \sigma}\equiv c_{i \sigma}^{\dag} c_{i \sigma}$.
 Hamiltonian (\ref{t-J exact complete}) in its complete form, i.e. with $c_1 = c_2 = 1$  is derived by applying canonical transformation to the  Hubbard Hamiltonian, $\hat{H}_{tU}$, in the strong coupling ($|t| << U $) limit, \cite{Spalek PRB, Spalek Acta} $\hat{H}_{tJ} = \exp({- i S}) \hat{H}_{tU}\exp({i S})$. Then, kinetic exchange integral $J_{ij} =  4|t_{ij}|^2/U$. On the other hand, the   $c_1 = c_2 = 0$ case has been used  as a simplified  effective model describing low-energy sector of a multi-band Hubbard model. \cite{Zhang Rice, Lee} Both the latter form, or that with  $c_1 = 1$, $c_2 =  0$,  are often used in studies of high- $T_{c}$ superconductivity. Our aim here is, among others, to compare the complete version of the $t$-$J$ model with the simpler forms. 
 
 It has been argued, \cite{Plain Vanilla} that a correct  variational   state describing the ground state of the $t$-$J$ Hamiltonian is  of the form   $| \Psi  \rangle  = \hat{P}  | \Psi_0 \rangle $, where  $| \Psi_0 \rangle$ is an uncorrelated state, here taken as the Bardeen-Cooper-Schrieffer  (BCS)-type  state.  When the Hubbard model is the starting point of analysis, the re-transformed back state with  perturbatively reintroduced double occupancies is used instead, i.e $| \Psi  \rangle \to |\tilde{\Psi}  \rangle = \exp{(i S)}| \Psi  \rangle $. \cite{Edegger Anderson Lett, Paramekanti} 
Also, the detailed form of   $| \Psi_0 \rangle $ may be   postulated without specifying the underlying microscopic Hamiltonian. \cite{Plain Vanilla}  
However,  within  RMFT,   $| \Psi_0 \rangle $  may  also  be selected  as an eigenstate of the effective single-particle   Hamiltonian,  obtained from the mean-field treatment of (\ref{t-J exact complete}). \cite{ZGRS, Edegger, The ladder of Sigrist, Ogata Himeda} This point of view is also taken up here.  

 Construction of  RMFT requires a prescription (here called the renormalization scheme, RS) for calculating expectation  value  of  an arbitrary operator $ \hat{\mathcal{O}} $ in the  correlated state  $ | \Psi  \rangle$. Explicitly,  
\begin{equation}
 \langle   \hat{\mathcal{O}} \rangle_{C} \equiv  \frac{\langle  \Psi |  \hat{\mathcal{O}}  | \Psi  \rangle}{\langle  \Psi    | \Psi  \rangle}   = \frac{\langle  \Psi_0 | \hat{P}_{C}  \hat{\mathcal{O}} \hat{P}_{C}  | \Psi_0  \rangle}{\langle  \Psi_0    | \hat{P}_{C}^{2} | \Psi_0  \rangle} \equiv  \frac{\langle   \hat{P}_{C}  \hat{\mathcal{O}} \hat{P}_{C}    \rangle}{\langle   \hat{P}_{C}^{2}   \rangle}.
\label{RS equation}
\end{equation} 
In the above, the projector $\hat{P}$  has been replaced by more general  correlator $\hat{P}_{C}$, which differs from $\hat{P}$  by the presence of fugacity factors,  ensuring that the  projected and  preprojected average particle numbers   are equal, i.e. $ \langle   \hat{n}_{i \sigma} \rangle_{C} = \langle   \hat{n}_{i \sigma}  \rangle$. \cite{Fukushima}


We adopt  the  RS of Ref. \onlinecite{Fukushima} and  supplement it with the variational formalism proposed recently by us. \cite{JJJS PRB, JJJS_arx_0} Combination of  these two factors yields  the results that differ considerably as compared to those of the standard RMFT. \cite{ZGRS, Edegger}
 To carry out the self-consistent variational approach, we start from the following effective MF grand Hamiltonian 
\begin{eqnarray}
\hat{K}_{\lambda}   &=&  - \sum_{\langle i j \rangle \sigma} \Big (\tilde{\eta}_{ij \sigma} \big (  c_{i \sigma}^{\dag} c_{j \sigma} - \chi_{ij \sigma}\big) + \text{H.c.} \Big)     \nonumber \\   &-&   \sum_{\langle i j \rangle  } \Big (\tilde{\gamma}_{ij} \big ( \hat{\Delta}_{ij} -  \Delta_{ij}\big) + \text{H.c.} \Big) - \mu \sum_{ i   \sigma}   \hat{n}_{i \sigma}    \nonumber \\   &-& \sum_{ i   \sigma} (\tilde{\lambda}^{n}_{i \sigma} \big ( \hat{n}_{i \sigma}  - n_{i \sigma}\big) \Big) + W(\chi_{ij \sigma}, \Delta_{ij}, n_{i\sigma}), 
\label{H no mu MF tJ like tilde}
\end{eqnarray} 
with $\hat{K}_{\lambda} \equiv \hat{H}_{\lambda} - \mu \hat{N}$, $W \equiv \langle  \hat{H}_{\text{tJ}} \rangle_{C} $, $n_{i\sigma} \equiv  \langle \hat{n}_{i \sigma}  \rangle$, $\chi_{ij} \equiv  \langle  c_{i \sigma}^{\dag}  c_{j \sigma} \rangle$, $\hat{\Delta}_{ij} \equiv  (c_{i \uparrow} c_{j \downarrow} - c_{i \downarrow} c_{j \uparrow})/2$, and   $ \Delta_{ij} \equiv  \langle \hat{\Delta}_{ij}\rangle$. The averages $ \langle \ldots  \rangle$ are defined by (\ref{RS equation}) for zero temperature, or in general, with the help of density operator $  \hat{\rho}_{\lambda}$, (see below). $\tilde{\eta}_{ij \sigma}$, $\tilde{\gamma}_{ij}$, and $\tilde{\lambda}^{n}_{i \sigma}$ (not to be confused with the fugacity factors of Ref. \onlinecite{Fukushima}) are the Lagrange multipliers, ensuring the self-consistency of this variational  approach, as this is not guaranteed within the standard Gutzwiller approximation. \cite{SGA}   
 The  form (\ref{H no mu MF tJ like tilde}),  apparently different from the   usual formulation of RMFT, is   fully equivalent to it (a similar approach is discussed in Ref. \onlinecite{Wang F C Zhang}).

We solve this model on a square lattice and in the spatially homogeneous situation, with no coexisting magnetic order. The model parameters are $t_1 \equiv t, t_3\equiv t^{\prime} $ and $t_5 \equiv  t^{\prime \prime} $, where $s= 1, 3, 5$ corresponds to   sites located at the distances  $d(i,j)$  of  $ 1, \sqrt{2}$, and $2$ lattice constants, respectively. Consequently, the following Lagrange multipliers and the corresponding mean fields are assumed as nonzero: $\chi_{ij} = \xi_s$ and $\eta_{ij} = \eta_s$;   $\Delta_{ij} = \Delta_{x(y)}$,   with $\Delta_x = \Delta = - \Delta_y$ and  $\tilde{\gamma}_{ij} =  \gamma_{x(y)} = \pm \gamma$, both for $d(i,j)=1$; $n_{i\sigma} = n_{i\bar{\sigma}}= n/2$  and $\tilde{\lambda}^{n}_{i \sigma} = \lambda$. Also, in (\ref{t-J exact complete}) we retain all the terms of the orders of $t^{2}/U$ and $tt^{\prime}/U$, and neglect a smaller term of the  order $(t^{\prime})^2/U$.  Then, $\langle  \hat{H}_{tJ} \rangle_{C} $ is obtained  using the  formalism of Ref. \onlinecite{Fukushima}. Explicitly,
\begin{equation}
W = \langle  \hat{H}_{t} \rangle_{C} + \langle  \hat{H}_{J} \rangle_{C} + \langle  \hat{H}_{3} \rangle_{C}\equiv W_{t} + W_{J} + W_{3},  
\label{W}
\end{equation}
where $W_{J} = \tilde{W}_{J} -\Lambda J c_1 n^2/2$, and $\Lambda$ is the number of lattice sites. In effect,  we have that
\begin{eqnarray} 
\frac{W_{t}}{\Lambda} & = &  
      \frac{ 16 (1-n)}{2-n} \Big\{ t_1 \xi _1  \left(1-\frac{4
  (\Delta ^2+\xi
   _1^2 )}{(2-n)^2} \right)  \nonumber \\ &+& \sum_{s =3, 5}  t_s \xi _s
    \left(1-\frac{4 \xi _s^2}{(2-n)^2} \right) \Big\},   
\label{W kin}
\end{eqnarray}

\begin{equation} 
\frac{\tilde{W}_{J}}{\Lambda}  =   -4J \left(  \frac{3
   \left(\Delta ^2+\xi _1^2\right) + c_1  (1-n)^2
   \left(\Delta ^2-\xi _1^2\right)}{(2-n)^2} \right). 
\label{W J}
\end{equation}
The formula for $W_{3}$, containing three-site terms, is too lengthy  to be reproduced here.

The next step is the diagonalization of $\hat{H}_{\lambda}$ (\ref{H no mu MF tJ like tilde}) via Bogoliubov-Valatin transformation, which leads to   
\begin{equation} 
\hat{K}_{\lambda} =  \sum_{\mathbf{k}} E_{\mathbf{k}} (\hat{\gamma}^{\dag}_{\mathbf{k}0}\hat{\gamma}_{\mathbf{k}0} + \hat{\gamma}^{\dag}_{\mathbf{k}1}\hat{\gamma}_{\mathbf{k}1}) + \sum_{\mathbf{k}} (\xi_{\mathbf{k}} - E_{\mathbf{k}}) + C,
\label{MF tJ BCS like}
\end{equation}
with $C = W + \Lambda (8\sum_{s} \xi_s \eta_s + 4\Delta \gamma + \lambda n)$. Also, the quasiparticle energy $E_{\mathbf{k}}$, the renormalized gap $D_{\mathbf{k}}$, and the renormalized band energy $\xi_{\mathbf{k}}$ are respectively given by
\begin{equation}
E_{\mathbf{k}} = \sqrt{\xi^2_{\mathbf{k}} + D^2_{\mathbf{k}}}, ~~~ D_{\mathbf{k}} = -\frac{\gamma}{2} \Gamma_{-}(\mathbf{k}),
\label{E_k D_k} 
\end{equation}
\begin{equation}
\xi_{\mathbf{k}} = -\left(\eta_1 \Gamma_{+}(\mathbf{k})  + \eta_3 \Theta(\mathbf{k}) +  \eta_5 \Gamma_{5}(\mathbf{k})  + \tilde{\mu}   \right), 
\label{xi_k} 
\end{equation}
$\Gamma_{\pm}(\mathbf{k}) = 2(\cos(k_x) \pm \cos(k_y))$,  $ \Theta(\mathbf{k}) = 4\cos(k_x) \cos(k_y)$, $\Gamma_{5}(\mathbf{k}) = 2(\cos(2 k_x) \pm \cos( 2 k_y))$, and $\tilde{\mu} \equiv \mu + \lambda$. 
Then, the corresponding grand-canonical density operator reads
\begin{equation}
\hat{\rho}_{\lambda} = \mathcal{Z}_{\lambda}^{-1}\exp(-\beta \hat{K}_{\lambda}), 
\label{rho moje} 
\end{equation}
with $\mathcal{Z}_{\lambda}=\text{Tr}[\exp(-\beta \hat{K}_{\lambda} )]$, $\beta = 1/k_B T$,  
and the generalized Landau functional,  $\mathcal{F} \equiv -\beta^{-1} \ln  \mathcal{Z}_{\lambda} $  reads
\begin{eqnarray}
\mathcal{F}  & = &C  + \sum_{\mathbf{k}}\big(  (\xi_{\mathbf{k}} - E_{\mathbf{k}}) - \frac{2}{\beta} \ln\big(1 + e^{-\beta E_{\mathbf{k}}}\big)\big).  
\label{mathcalF tJ BCS like BCS}  
\end{eqnarray}
 Because of extra constraints in (\ref{H no mu MF tJ like tilde}), the  necessary conditions for  $\mathcal{F}$ to have a minimum are 
\begin{eqnarray}
\partial_{w} \mathcal{F} = 0, ~~~  \partial_{z} \mathcal{F} = 0, 
\label{derivative of mathcalF A, lambda}  
\end{eqnarray}
for the  mean-fields $w \in \{\xi_s, \Delta, n \}$,  $n \equiv 1 - x$, and the Lagrange multipliers $z \in \{\eta_s, \gamma, \lambda\}$.
In contrast to our previous formulation, \cite{JJJS PRB}  the  equations   $\partial_{w} \mathcal{F} = 0$   are   solved analytically,  but the solution of  those remaining   ($\partial_{z} \mathcal{F} = 0$) must be determined numerically. 
Optimal values of the mean-fields and the Lagrange multipliers, given by a solution of (\ref{derivative of mathcalF A, lambda}) characterized with the lowest value of $\mathcal{F}$ and denoted respectively $(\xi^{(0)}_{1}, \xi^{(0)}_{3}, \xi^{(0)}_{5}, \Delta^{(0)}, n) \equiv \vec{A}_0(T, V, \mu)$, $(\eta^{(0)}_{1}, \eta^{(0)}_{3}, \eta^{(0)}_{5}, \gamma^{(0)}, \lambda^{(0)}) \equiv \vec{\lambda}_0(T, V, \mu)$, may be used next to construct standard grand thermodynamic potential $\Omega$ and the free energy $F = \Omega + \mu N$. Explicitly,
\begin{equation}
\Omega(T, V, \mu) = \mathcal{F}(T, V, \mu; \vec{A}_0(T, V, \mu),  \vec{\lambda}_0(T, V, \mu)).
\label{Thermo potentials bis}
\end{equation}
The present formalism, based on the maximum-entropy principle \cite{Jaynes} is formally valid for arbitrary $T>0$. Consequently, we have replaced pure state  $|\psi_0 \rangle $ by mixed state represented by $ \hat{\rho}_{\lambda}$. However, the solutions obtained for non-zero, but sufficiently low $T$, are for all practical purposes identical to those for $T=0$. This situation is studied in the next Section. The separate question is that of the validity of RMFT approach at $T>0$ from the point of view of physical consistency. This is analyzed in more detail in Section IV. 

\section{\label{sec:3} Results for $T\approx 0$}
An analytical expression for $W = \langle  \hat{H}_{tJ} \rangle_{C}$ allows us to make qualitative predictions before the numerical analysis is carried out.  
First, from (\ref{W kin}) we expect a strong tendency to the  superconductivity suppression for  higher doping, as SC order leads to the band energy decrease $\sim \Delta^2$. On the other hand, in that regime the renormalized band energy becomes predominant over the exchange part. In effect, the normal state is    favored over SC for   $x > x_c$ with $x_c$ smaller than obtained within previous MF treatments. \cite{ZGRS, Edegger, Plain Vanilla, Edegger Anderson Lett, Shih Lett} Second, from (\ref{W J}) we infer, that the influence of the $ \hat{\nu}_{i}\hat{\nu}_{j}/4$ term  on $W_J$ is small except for the  largest doping; this is due to the presence of the $(1-n)^2$  prefactor (the other term $\sim n^2$ merely shifts the chemical potential). 
On the other hand, $W_3$ is multiplied only by $(1-n) = x$ prefactor. Consequently,  for higher $x$ this term becomes rather important,  due to  the number of distinct three-site terms   present for a given initial site and spin direction (eight for $d(i,j) = \sqrt{2}$, and four for $d(i,j) =  2$). Also, this part  of $W$ is expected to suppress SC order, as the term $\sim\Delta^2$ in $W_3$ contains a factor,  which is positive for  reasonable values of other mean fields and model parameters. 
 
\begin{figure}[h]
\begin{center}
\rotatebox{270}{\scalebox{0.27}{\includegraphics{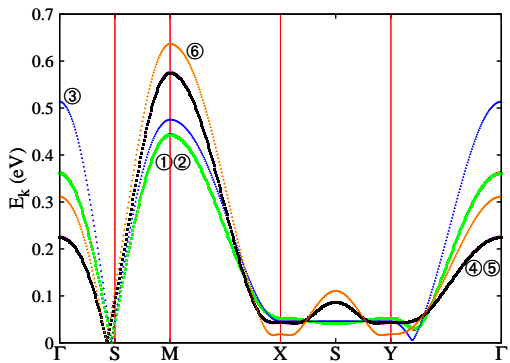}}}
\end{center}
\caption{(Color online) Dispersion relations along the main symmetry lines in the two-dimensional Brillouin zone for $n=0.825$. The various curves are explained in main text. Note, the curves 1 and 2, as well as 4 and 5 are practically indistinguishable (the influence of the term $\sim c_1$ in (\ref{t-J exact complete}) is neglible).}
\label{dispertion for 256 fuk RVB}
\end{figure}

\begin{figure}[h!]
\begin{center}
\rotatebox{270}{\scalebox{0.27}{\includegraphics{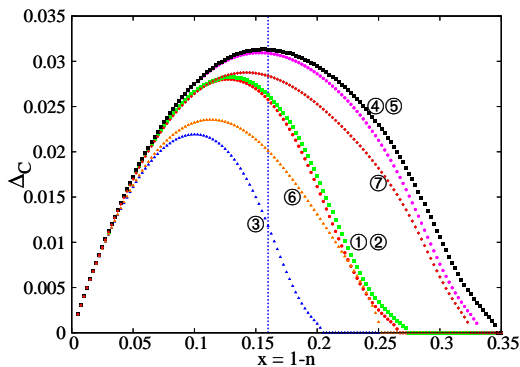}}}
\end{center}  
\caption{(Color online) Doping dependence of the renormalized superconducting order parameter  $\langle \hat{\Delta} \rangle_C \equiv \Delta_C$. The curves 1-6 correspond to those in Fig. 1. The curve 7 is for $c_1 =1$, $c_2 =0$,  $t^{\prime}/t = -0.27$, $t^{\prime \prime} = 0$, and  for $J/|t| =0.3$. For definitions of cases 1-7 see the main text. Cases 3 and 6 correspond to the complete $t$-$J$ Hamiltonian (\ref{t-J exact complete}) with $c_1=c_2$.}
\label{x dependence of ren delta}
\end{figure}

\begin{figure}[h!]
\begin{center}  
\rotatebox{270}{\scalebox{0.32}{\includegraphics{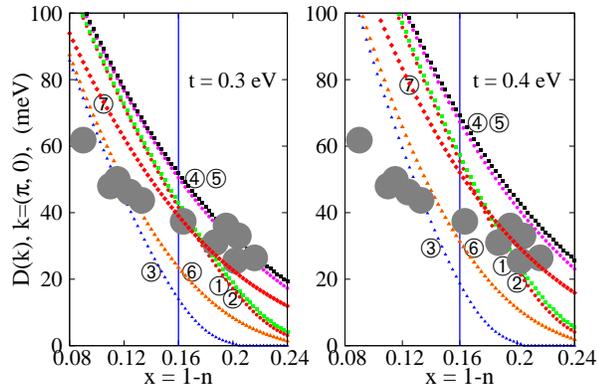}}}
\end{center}
\caption{(Color online) Doping dependences of the  SC gap $D_{\mathbf{k}}$ at $\mathbf{k} = (\pi, 0)$ for  cases 1-6 and for $t^{\prime}/t = -0.27$, and $J/|t| =0.3$ (filled diamonds). Large filled circles - experimental data. \cite{Campuzanno}  Note, that in contradistinction to Ref. \cite{Edegger Anderson Lett} no \textit{ad hoc} introduced scaling factor $\alpha = 1/2$ is necessary to obtain a reasonable agreement in the overdoped regime, i.e. to the right of the vertical line.}
\label{doping dependance of D_k zero pi}
\end{figure}

\begin{figure}[h!] 
\begin{center} 					 
\rotatebox{270}{\scalebox{0.32}{\includegraphics{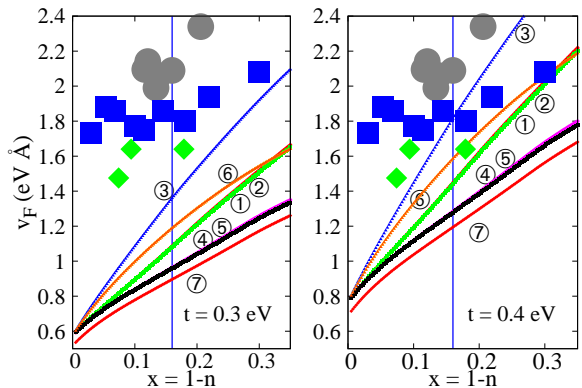}}} 
\end{center} 
\caption{(Color online) Doping dependence of Fermi velocity in the nodal ($  (0, 0) \to (\pi, \pi)$) direction. Experimental data (cf. \cite{Edegger Anderson Lett} and References therein) are marked by diamonds (YBCO), squares (LSCO)  and solid circles (BSCCO).} 
\label{doping dependance v_F} 
\end{figure}

 
We solve numerically the part $\partial_{z} \mathcal{F} = 0$ of Eqs. (\ref{derivative of mathcalF A, lambda}) for the mean fields using   periodic boundary conditions on the  lattice of  $ \Lambda = 512^2 $ sites, to minimize finite size effects  and for  low temperature   $k_B T  = 2 \cdot 10^{-3}  J$. The solution amounts to solving simultaneously the system of five nonlinear equations using GNU Scientific Library (GSL). In the  most cases we take the parameters $ |t|/J = 3$ (corresponding to $U/|t| = 12$ for the Hubbard model), $ t^{\prime}/t = 0$ or $(-0.25)$, and $t^{\prime \prime} = 0$. Additionally, we take also $|t|= 0.3$ eV or $|t|= 0.4$ eV, which correspond roughly to the lower and the upper limits of  the realistic values of this parameter, depending on the compound. Values of $|t|$ close to  $0.4$ eV have been determined from the band-structure calculations, \cite{Lee, Hybertsen} whereas $|t|= 0.3$ eV is used in Refs. \onlinecite{Edegger Anderson Lett, Paramekanti}. To highlight the influence of various forms of (\ref{t-J exact complete}),  the results for   different values of $c_1$ and $ c_2$ are analyzed. The  numbers 1, 2, 3 (4, 5, 6) in Figs. \ref{dispertion for 256 fuk RVB}-\ref{doping dependance v_F} and Table I correspond to the three situations:  $c_1 = c_2 = 0$ (i.e. with the $\hat{\mathbf{S}}_{i}\cdot \hat{\mathbf{S}}_{j}$ part of the kinetic exchange only, cf. (\ref{t-J exact complete})), $c_1 = 1$ and $c_2 = 0$ (with full form of the kinetic exchange),  and $c_1 = c_2 = 1$ (complete form of the $t$-$J$ Hamiltonian with the three-site terms included), each case taken for $ t^{\prime}/t = 0$ ($ t^{\prime}/t =-0.25$), respectively. 

In Table I we provide the equilibrium values of the mean-fields and of the Lagrange multipliers for cases 1-6 and for $x= 0.175$, a representative hole concentration in the overdoped regime.
\begin{center}
Table I. Optimal values of mean-field parameters for $T\approx 0$ and $x= 0.175$.\\
\begin{tabular}{c | c c c | c  c c}\hline \hline
$\varphi$   		& 1       &  2     &  3     & 4 	 &  5    &  6  \\  \hline 
$ \xi_1 $   		& 0.1970  & 0.1969 & 0.1990 & 0.1924 & 0.1922  & 0.1944          \\  
$ \xi_3 $   		& 0.0468  & 0.0465 & 0.0505 & 0.0241 & 0.0239  & 0.0225     \\  
$ \xi_5 $   		& -0.0080 &-0.0076 &-0.0144 & 0.0337 & 0.0340  & 0.0383      \\  
$\Delta_x$  		& 0.0687  & 0.0708 & 0.0202 & 0.0903 & 0.0919  & 0.0534   \\  \hline
$\eta_1 $   		& 1.0080  & 1.0030 & 1.2355 & 1.0031 & 0.9982  & 1.1845       \\ 
$\eta_3 $   		& 0.0000  & 0.0000 & 0.0803 &-0.2223 &-0.2223  &-0.2118       \\   
$\eta_5 $   		& 0.0000  & 0.0000 & 0.0408 & 0.0000 & 0.0000  & 0.0404       \\   
$\tilde{\gamma}_x $     &  0.1584 & 0.1665 & 0.0320 & 0.2126 & 0.2205  & 0.0834  \\   
$\tilde{\mu}  $ 	& -0.4069 &-0.4080 &-0.2935 &-0.8633 &-0.8614  &-0.9406    \\   
\hline \hline
\end{tabular}
\end{center}
In Fig. \ref{dispertion for 256 fuk RVB} we plot the dispersion relation for the Bogoliubov quasiparticles, calculated for the parameters displayed in Table I. The influence of $\hat{H}_{3}$ on $E_{\mathbf{k}}$ is of  comparable magnitude to that of having nonzero $t^{\prime}$. 

Next, we discuss the doping-dependence of the renormalized SC order parameter $\langle \hat{\Delta}_{ij} \rangle_{C} \equiv \Delta_C  $ (cf. Eqn. (18) of Ref. \onlinecite{Fukushima}).
 The numerical results confirm  the above made qualitative predictions. In Fig. \ref{x dependence of ren delta} we plot $\Delta_C$ for the cases 1-6 specified above, as well as for $t^{\prime}/t = -0.27$ (value being reasonable for BSCCO compounds, \cite{Pavarini}) and  $J/|t| =0.3$  (curve 7). 
Note, that the upper  critical concentration $x_c$ for the  cases 4 and 5 is close to  the VMC result, \cite{Paramekanti} obtained within the Hubbard model and using $\tilde{| \psi \rangle}$ for the corresponding model parameters.  
Also, nonzero value of $t^{\prime}$ enhances superconductivity, in agreement with previous VMC results  \cite{Shih Lett} and other calculations.\cite{Pavarini} Let us emphasize again, that the presence of the $\hat{H}_{3}$ term acts in the opposite direction. The vertical line roughly marks the boundary between under- and over-doped regimes. Importantly, for $|t|/J = 3$, $t^{\prime \prime} = 0$ and  different   $t^{\prime}/t$ values, $0 \leq t^{\prime}/t \leq 0.25$, $x_c$ lies in the interval $0.2 \lesssim x_c \lesssim  0.35$, depending on the form of $\hat{H}_{\text{tJ}}$, as illustrated in Fig. \ref{x dependence of ren delta}.  
As said above, the small difference between the curves 4 and 5 (as well as between 1 and 2) shows an insignificane of the term $\sim c_1$ \textsf{(cf. also Table I)}.  
The results for $x_c$ are in a good overall agreement with the  experimental data for the cuprates.\cite{Lee, Nakano exp} 
To provide an additional support for  our results, we list in Table II the values of  $x_c$ as a function of $J$,  for either $t^{\prime}/t = -0.1$ (considered to be relevant to the LSCO compound\cite{Lee}), or  $t^{\prime}/t = -0.27$.

A remark is in place here. The family of curves in Fig. \ref{x dependence of ren delta} and of $x_c$ values in Table II has the following meaning. Each $x_c$ value singles out either the choice of a model or a particular set of parameters. This detailed analysis is to illustrate that there is a clear upper critical concentration in the proper range, irrespectively of the model details or particular set of parameter values.
\begin{center}
Table II. Upper critical concentration $x_c$ vs. $J$ for  $t^{\prime} = - 0.27t$ or $t^{\prime} = - 0.1t$. The symbol A (B) labels the case $c_2=0$ and  $c_1 = 0$ ($c_1 = 1$), respectively, whereas C means that $c_1 = c_2 = 1$ is taken in the computation.

\vspace{0.1cm}
\begin{tabular}{c|c c c c c}\hline \hline
 $J/|t| $ & 0.2  &  0.3  &  0.333    &   0.375    & 0.4             \\  \hline 
$t^{\prime}/t = -0.1$ A   &  0.18       &   0.26      &    0.29  &  0.32  &  0.33         \\   
$t^{\prime}/t = -0.27$ A  & 0.2   &   0.31        &  0.34   &  0.38  &     0.4   \\     \hline  
$t^{\prime}/t = -0.1$ B  &  0.18       &  0.27       &  0.3    &  0.33     &   0.35         \\   
$t^{\prime}/t = -0.27$ B  & 0.2     &    0.33       &  0.36   &  0.4   &   0.42   \\     \hline  
$t^{\prime}/t = -0.1$ C  &  0.15        &      0.21    &   0.22   &  0.24     &  0.25        \\   
$t^{\prime}/t = -0.27$ C  &   0.15   &   0.23       &  0.26     &  0.28   &    0.29     \\  
\hline \hline
\end{tabular}
\end{center}

In Fig. \ref{doping dependance of D_k zero pi} we plot  $x$-dependence of the SC gap $D_{\mathbf{k}}$ for $\mathbf{k}=(\pi, 0)$ (c.f. Eqn. (\ref{E_k D_k})) and compare our results with the experimental data.\cite{Campuzanno}
For the  selections  of  $ t^{\prime}/t$ and $J/|t|$ as in Fig. \ref{dispertion for 256 fuk RVB}, no fully    satisfactory agreement with  experiment  is achieved in the entire range of $x$. However, the   agreement  with experiment is quite good for the parameters corresponding to the curves 1, 2, 4 and 5  in the overdoped regime, both for $|t|= 0.3$ eV and $|t|= 0.4$ eV. The best overall fit is achieved for the set of parameters represented by curve 7.
Note, that  in all the cases 1-7 the quasiparticle energies  obtained here are decisively  lower than those in the standard RMFT formulation. \cite{Edegger Anderson Lett} These differences are caused  by both the  particular selection of the renormalization scheme, as well as by the   variational method we use. As a consequence, we obtain also lower values of the Fermi velocity, $v_F \equiv |\nabla_k \xi_{\mathbf{k}}|_{|\mathbf{k}| = k_F}$, calculated  for the nodal ($(0,0)\to(\pi, \pi)$) direction. The lattice constant has been taken as  $a_0 = 4\text{\r{A}}$. \cite{Edegger Anderson Lett} 

The $x$-dependence of  $v_F$ is detailed in Fig.  \ref{doping dependance v_F} for the same set  of parameters as in Figs. \ref{x dependence of ren delta} and \ref{doping dependance of D_k zero pi}, for both  $|t|= 0.3$ eV  and $|t|= 0.4$ eV, and compared  with the data discussed before. \cite{Edegger Anderson Lett} The theoretical values are still too low. Also, the $x$-dependence of both $D_{\mathbf{k} =(\pi, 0)}$ and $v_F$, obtained within the MF approaches, is stronger than observed in experiment. This feature is shared with the other mean-field approaches. \cite{Edegger Anderson Lett, Paramekanti}
However,   the experimental values for BSCCO   $\sim$ 1.5 eV $\text{\r{A}}$ have also been reported, \cite{Kaminski nice} and are quite close to our results. Note that the disagreement is largest in the underdoped regime (to the left of the vertical line). 

\section{\label{sec:4} Results for $T > 0$.} 
So far, we have analyzed only low temperature $\beta J =k_B T/J = 500$, practically equivalent to the true $T=0$ situation. Obviously, it would be interesting to extend the analysis for higher temperatures, and in particular, to determine the critical temperature $T_c$ as a function of doping. Standard RMFT approach, as based on the Gutzwiller approximation (GA), is devised to study ground-state properties and as such, is not applicable directly for $T>0$. It may seem, that having a finite-temperature formalism at our disposal, we may examine arbitrary temperature by simply changing value of $\beta$ in $\mathcal{F}(\beta)$ Eq. (\ref{mathcalF tJ BCS like BCS}). Unfortunately, this may lead to the situation similar to that encountered in  the slave-boson mean-field theories, which application at finite temperatures is invalidated by incorrect evaluation of the entropy part of the free energy. \cite{Gebhard} 


Recently, an attempt to extend RMFT to $T>0$ have been made. \cite{TRMFT} Within this approach, termed \textit{finite temperature RMFT} (TRMFT), the term 
\begin{eqnarray}
\Delta S = - \sum_{i} \left(e_i \ln \frac{e_i}{e_{i0}} + q_i \ln \frac{q_i}{q_{i0}} + d_i \ln \frac{d_i}{d_{i0}} \right). 
\label{Delta entropy of Wang}  
\end{eqnarray} 
is added to the single-particle entropy $S_0 = -\text{Tr}\hat{\rho}_0 \ln \hat{\rho}_0$ of the mean-field model (here, $\hat{\rho}_0 = \hat{\rho}_{\lambda}$). In Eq. (\ref{Delta entropy of Wang}), we have $e_i = \langle \hat{E}_i \rangle_C$, $e_{i0} = \langle \hat{E}_i \rangle$, $q_i = \langle \hat{Q}_i \rangle_C$, $q_{i0} = \langle \hat{Q}_i \rangle$, $d_i = \langle \hat{D}_i \rangle_C$, and $d_{i0} = \langle \hat{D}_i \rangle$ with $\hat{E}_i = (1-\hat{n}_{i\uparrow})(1-\hat{n}_{i\downarrow})$, $\hat{Q}_i = \hat{n}_{i\uparrow}(1-\hat{n}_{i\downarrow}) + \hat{n}_{i\downarrow}(1-\hat{n}_{i\uparrow})$ and  $\hat{D}_i = \hat{n}_{i\uparrow}\hat{n}_{i\downarrow}$. Note, that $\Delta S < 0$. Derivation of (\ref{Delta entropy of Wang}), and its possible generalizations will be discussed elsewhere. \cite{Our TRMFT} Here we simply adapt Eq. (\ref{Delta entropy of Wang}), which may be treated as a reasonable \textit{Ansatz}, similar in spirit to the finite-temperature extensions of the Gutzwiller approximation proposed earlier. \cite{Szef T 1, Seiler, Szef T 2}

Within the present formalism, $\Delta S$ can be included by replacing $W(\xi_s, \Delta, n)$ (\ref{W}) by $W - T\Delta S$. However, for the $t$-$J$ model, $d_{i} = 0$, and consequently, for non-magnetic, homogeneous solutions studied here, $\Delta S \equiv \Delta S_{tJ}$ depends only on the total particle number $n$, i.e.
\begin{equation}
\frac{\Delta S_{tJ}}{\Lambda} =  (2-n) \ln\left(1-\frac{n}{2}\right) - (1-n) \ln(1-n).
\label{Delta entropy tJ}
\end{equation}
Therefore, the presence of $\Delta S_{tJ}$ results only in different values of $\mu$, $\lambda$, and thermodynamic potentials $\Omega$ and $F$ (\ref{Thermo potentials bis}). Still, $\tilde{\mu}=\mu+\lambda$, all the mean fields, and the remaining Lagrange multipliers, as well as the free energy difference between the superconducting and the normal solutions, remain unchanged. Hence, $\Delta S$ in the form (\ref{Delta entropy of Wang}) does not lead to any nontrivial modifications of our original formulation.

\begin{figure}[h!] 
\begin{center} 					 
\rotatebox{270}{\scalebox{0.32}{\includegraphics{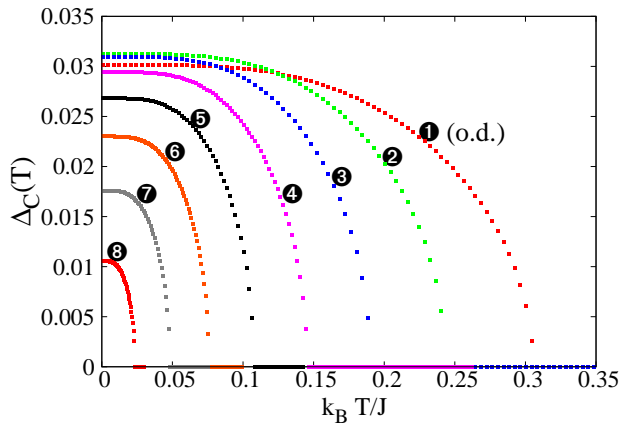}}} 
\end{center} 
\caption{(Color online) Temperature dependence of renormalized superconducting order parameter  $\langle \hat{\Delta} \rangle_C(T) \equiv \Delta_C(T)$. The numbers 1-8 correspond to values of doping equal $0.125$ (optimal doping, o.d.), $0.15$, $0.175$, $0.2$, $0.225$, $0.25$, $0.275$ and $0.3$, all for $c_1 =1$, $c_2 =0$,  $t^{\prime}/t = -0.25$.} 
\label{T dependance ren Delta} 
\end{figure}

Nonetheless, it still seems to be interesting to apply the above formalism to examine nonzero temperature situation. This would reveal limitations of the present form of RMFT and should help in formulating more satisfactory finite-temperature mean-field treatment of the $t$-$J$ model. First, in Fig. \ref{T dependance ren Delta} we plot temperature dependence of (renormalized) gap magnitude $\Delta_C(T)$ as a function of $T$ for selected hole concentrations. 
For the values of model parameters used here, we have $J = 100 \text{meV} \approx 1160 \text{K}$ ($J = 133 \text{meV}  \approx 1550 \text{K} $) for $t = 300 \text{meV}$ ($t = 400  \text{meV}$), respectively. Therefore, the critical temperature is overestimated by a factor 3-5 (e.g. $T_c = 340 - 450 \text{K}$ at the optimal doping and  $T_c \approx 25 - 35  \text{K}$ at $x=0.3$. This is a common feature of all mean-field type approaches, clearly caused by neglecting the fluctuations. This also shows the insufficiency of the expression $\Delta S$ (\ref{Delta entropy of Wang}) in the present $t$-$J$ model case.

\begin{figure}[h!] 
\begin{center} 					 
\rotatebox{270}{\scalebox{0.32}{\includegraphics{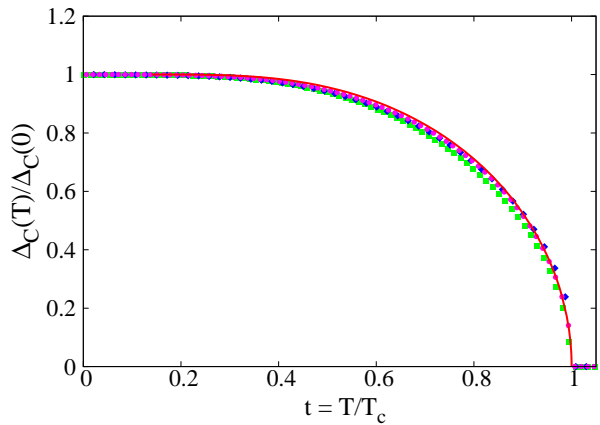}}} 
\end{center} 
\caption{(Color online) Reduced temperature dependence of renormalized superconducting order parameter  $\langle \hat{\Delta} \rangle_C \equiv \Delta_C(T)$ for various doping levels. Squares: $x=0.125$, solid circles: $x=0.25$, diamonds: $x=0.3$ are all for $c_1 =1$, $c_2 =0$,  $t^{\prime}/t = -0.25$. Solid line: BCS result.} 
\label{BCS behavior} 
\end{figure}

Nonetheless, one interesting property of the present approach should be noted. Namely, an uncorrelated wave function $| \Psi_0  \rangle$, or a related density operator $\hat{\rho}_{\lambda}$ (\ref{rho moje}) has an essentially the same form as that coming from the Bardeen-Cooper-Schrieffer (BCS) theory. Therefore, it seems natural to compare our values of reduced renormalized gap magnitude $\Delta_C(T)/\Delta_C(0)$ with the standard BCS result \cite{Rickayzen} given by $\Delta(T)/\Delta(0) = \tanh (\Delta(T)/t\Delta(0))$, where $t=T/T_C$. The results are shown in Fig. \ref{BCS behavior}.  
Interestingly, the agreement with the BCS results is quite good, despite the renormalization of $\Delta$ and its nontrivial d-wave symmetry. Uncorrelated (bare) gap ratio $\Delta(T)/\Delta(0)$ exhibits very similar scaling with $T/T_c$.  

\section{\label{sec:5} Summary  and outlook} 
In this paper, we aimed to describe the doping dependences of selected, experimentally measured quantities, we have decided to systematize the results coming from different versions of $t$-$J$ model, that are discussed in the literature. From that analysis it follows that while the presence of the term $\sim c_1$ does not influence remarkably the results except for the largest doping $x$, the inclusion of the three-site terms ($c_1=c_2=1$) reduces substantially the range in which superconductivity is present.

We have implemented the mean-field renormalization scheme \cite{Fukushima} to $t$-$J$ model, in which both the kinetic-exchange and the three-site terms can be taken into account. Such   RMFT approach,  which is based on an effective single-particle picture, is  expected to be valid  first of all  in the overdoped regime, where an unconventional form of the Fermi liquid is obtained. 
The variational approach based on the maximum entropy principle \cite{Jaynes} has been used to determine mean-field parameters appearing in the model. The theoretical results  yield a correct value for the upper critical concentration for the high-$T_c$ d-wave superconductivity disappearance. 

Our method provides also lower single-particle energies compared to the previous MF results. \cite{Edegger Anderson Lett} Consequently, a good estimate of the experimentally determined gap $D_{\mathbf{k} =(\pi, 0)}$ is detected experimentally in the overdoped regime for typical values of the model parameters. However, the values of the Fermi velocity in the nodal direction are still slightly too low. We also have examined the temperature dependence of the superconducting gap magnitude and have determined the critical temperature $T_c(x)$ evolution as a function of hole concentration. As may be expected, mean-field results overestimate $T_c$ by a factor 3-5. The present results can be generalized by taking into account more complex lattice or band structure, the broken $C_{4v}$ symmetry (Pomeranchuk instability \cite{JJJS PRB}), and  antiferromagnetism. Furthermore, study of the Fermi-surface-topology evolution with the doping is achievable. Finally, a more advanced scheme of calculating the entropy of the correlated state is desired to provide more satisfactory description of the nonzero-temperature situation. We should be able to see progress along these lines soon.

\section{\label{sec:6} Acknowledgments} 
We thank Prof. Maciej Ma\'{s}ka for discussions.
The technical help of Marcin Abram, Andrzej Biborski, Jan Kaczmarczyk, Andrzej Kapanowski, Micha\l ~K\l os, and Magdalena Koz\l owska is warmly acknowledged.
The work was supported by  the Grant No. N N 202 128 736 from the Ministry of Science and Higher Education.

\end{document}